# Unconventional gapping behavior in a kagome superconductor

Md Shafayat Hossain[1*†], Qi Zhang[1*], Eun Sang Choi[2*], Danilo Ratkovski[2*], Bernhard Lüscher[3*], Yongkai Li[4,5,6*], Yu-Xiao Jiang[1], Maksim Litskevich[1], Zi-Jia Cheng[1], Jia-Xin Yin[1], Tyler A. Cochran[1], Brian Casas[2], Byunghoon Kim[1], Xian Yang[1], Jinjin Liu[4,5,6], Yugui Yao[4,5], Ali Bangura[2], Zhiwei Wang[4,5,6†], Mark H. Fischer[3], Titus Neupert[3†], Luis Balicas[2†], M. Zahid Hasan[1,7†]

**Affiliations:**

[1]Laboratory for Topological Quantum Matter and Advanced Spectroscopy, Department of Physics, Princeton University, Princeton, New Jersey, USA.

[2]National High Magnetic Field Laboratory, Tallahassee, Florida 32310, USA.

[3]Department of Physics, University of Zurich, Winterthurerstrasse, Zurich, Switzerland.

[4]Centre for Quantum Physics, Key Laboratory of Advanced Optoelectronic, Quantum Architecture and Measurement (MOE), School of Physics, Beijing Institute of Technology, Beijing 100081, China.

[5]Beijing Key Lab of Nanophotonics and Ultrafine Optoelectronic Systems, Beijing Institute of Technology, Beijing 100081, China.

[6]Material Science Center, Yangtze Delta Region Academy of Beijing Institute of Technology, Jiaxing 314011, China.

[7]Quantum Science Center at ORNL, Oak Ridge, TN, USA.

†Corresponding authors, E-mail: mdsh@princeton.edu; zhiweiwang@bit.edu.cn; titus.neupert@physik.uzh.ch; balicas@magnet.fsu.edu; mzhasan@princeton.edu.

*These authors contributed equally to this work.

## Abstract
**Determining the types of superconducting order in quantum materials is a challenge, especially when multiple degrees of freedom, such as bands or orbitals, contribute to the fermiology and when superconductivity competes, intertwines, or coexists with other symmetry-breaking orders. Here, we study the Kagome-lattice superconductor $CsV_3Sb_5$, in which multiband superconductivity coexists with a charge order that substantially reduces the compound's space group symmetries. Through a combination of thermodynamic as well as electrical and thermal transport measurements, we uncover two superconducting regimes with distinct transport and thermodynamic characteristics, while finding no evidence for a phase transition separating them. Thermodynamic measurements reveal substantial quasiparticle weight in a high-temperature regime. At lower temperatures, this weight is removed via the formation of a second gap. The two regimes are sharply distinguished by a pronounced enhancement of the upper critical field at low temperatures and by a switch in the anisotropy of the longitudinal thermal conductivity as a function of in-**

**plane magnetic field orientation. We argue that the band with a gap opening at lower temperatures continues to host low-energy quasiparticles, possibly due to a nodal structure of the gap. Taken together, our results present evidence for band-selective superconductivity with remarkable decoupling of the (two) superconducting gaps. The commonly employed multiband scenario, whereby superconductivity emerges in a primary band and is then induced in other bands appears to fail in this unconventional kagome superconductor. Instead, band-selective superconducting pairing is a paradigm that seems to unify seemingly contradicting results in this intensely studied family of materials and beyond.**

## Main text

Superconductivity in a system with multiple Fermi surfaces involving Cooper pairing has attracted significant interest due to the two-gap superconductivity of $MgB_2$ or the Fe-based superconductors. Multiple superconducting gaps lead, for example, to remarkably high superconducting upper critical fields[1], temperature-dependent superconducting anisotropies[2], and total superfluid densities that can only be described by the sum of contributions from the different bands involved in the superconducting pairing[3]. Models treating these systems assume a predominant superconducting gap that induces superconductivity in other bands via the superconducting proximity effect due to effective inter-band interactions. This leads to the simultaneous opening of superconducting gaps on multiple bands and therefore to a single anomaly in thermodynamic measurements, such as the heat capacity $C/T$, at the superconducting transition temperature $T_c$[4]. Nevertheless, in $MgB_2$ the additional SC gap leads to a broad and shallow anomaly in $C/T$ below $T_c$[5].

In this context, the new family of superconducting kagome compounds, crystallizing in the hexagonal $P6/mmm$ space group and belonging to the generic family $AV_3Sb_5$ (where $A$ = K, Rb, Cs)[6], offers a unique platform to study multiband superconductivity emerging within a (competing) phase. The electronic band structure, measured by angle-resolved photoemission spectroscopy, is characterized by several bands crossing the Fermi level, which stem from the V $d$ orbitals and the Sb $p$ orbitals, and, according to first-principles calculations, displays non-trivial $Z_2$ topological invariants at several direct band gaps[7]. Remarkably, superconductivity nucleates in a metallic state that displays a $2 \times 2 \times 2$ (or $2 \times 2 \times 4$) charge order at $T_{CO}$ = 94 K, which was suggested to break time-reversal symmetry at lower temperatures[8,9]. Still, some experiments suggest that the superconductivity in $CsV_3Sb_5$ might be of conventional nature. Knight-shift experiments point towards a spin-singlet pairing configuration[10], which is supported by the observation of a Hebel-Slichter peak in $1/T_1T$[10]. Furthermore, a full gap without nodes was also reported via magnetic penetration depth experiments[11,12]. The picture is, however, complicated by the observation of a residual density of states in the superconducting state[13,14] and the potential time-reversal-symmetry-breaking background of the charge density wave phase[8,9]. Although evidence for time-reversal symmetry down to low temperatures, where the superconducting state condenses, remains controversial, current experiments indicate that the charge order breaks the lattice rotation symmetry, from the six-fold inherent to the kagome lattice to two-fold[15,16]. A two-fold symmetric upper critical field was also observed as the magnetic field is rotated within the planes, implying that the superconducting state itself acquires a nematic character[15].

Several studies on the superconducting ground state point to a nodeless but multi-gap superconducting state[10,11,17], albeit some of these reports favor a time-reversal-symmetry-broken state[17,18]. Coexistence with a nematic charge order which seemingly breaks time-reversal symmetry, and a multi-orbital character, could even be expected to conspire to produce an unconventional superconducting state whose nature remains to be unveiled[12,13,14,17,18,20,21,22].

Here, we explore the superconducting phase diagram of CsV$_3$Sb$_5$, via electrical and thermal transport experiments, as well as heat capacity measurements, particularly for fields along a planar direction, finding a superconducting phase diagram that seems to be composed of two distinct superconducting regimes. The sharp distinction between both regimes under a magnetic field and at zero field, not obviously seen in other multiband superconductors such as MgB$_2$ or the Fe pnictides, suggests a remarkable decoupling of the involved degrees of freedom potentially involving an orbital-selective pairing[23] as observed in FeSe[24].

We combine several experimental techniques, namely electrical and thermal transport, as well as specific heat measurements, to investigate the superconducting phase diagram of CsV$_3$Sb$_5$. For the electrical transport measurements, we fabricated four-point probe samples (as shown in the inset of Fig. 1**b**) using mechanically exfoliated crystals with a thickness of approximately 90 nm (see Methods Section II and Extended Fig. 1), which is sufficiently thick to capture the bulk properties of CsV$_3$Sb$_5$. Details on the crystal growth and sample fabrication can be found in Methods Sections I and II. Upon cooling the exfoliated CsV$_3$Sb$_5$ sample, the resistivity of the sample gradually decreases, indicating metallic behavior, until it undergoes a transition to a superconducting state with a critical temperature (defined as the temperature, where the resistivity drops to 50% of its original value in the normal state just above $T_c$) of $T_c = 3.5$ K (Fig. 1**b**).

To explore the characteristics of this superconducting state, we examined the magnetoresistivity of CsV$_3$Sb$_5$ (Fig. 1**c**) at very low temperatures ($T = 0.02$ K) as a function of magnetic field orientation, ranging from the *c*-axis to the *ab*-plane. An important piece of information obtained from the magnetoresistivity traces, shown in Fig. 1**c**, is the anisotropy of the upper critical magnetic field ($\mu_0 H_{c2}$), defined as the magnetic field at which the resistivity drops to 50% of its normal state value. In Fig. 1**d**, we plot $\mu_0 H_{c2}(\theta)$ as a function of the angle between the *c*-axis and the magnetic field orientation. As the magnetic field direction is rotated from the out-of-plane ($\theta = 0^o$) to an in-plane ($\theta = 90^o$) orientation with respect to the *ab*-plane of CsV$_3$Sb$_5$, the transition to the (field-induced) normal state progressively shifts to higher fields, with the highest $\mu_0 H_{c2}(\theta)$ occurring at $\theta = 90^o$ (Fig. 1**c**, **d**). This observation indicates a clear superconducting anisotropy, which, in our case, can be best fit to the Tinkham formula[25] describing the angular dependence of $\mu_0 H_{c2}(\theta)$ for a two-dimensional superconductor (see Methods Section IV). While our data can be well fitted to the Tinkham formula as depicted in Fig. 1**d**, the fitting shows discrepancies (particularly for large $\theta$ values) with the Ginzburg-Landau anisotropic mass model[24], which describes the angular dependence of $H_{c2}$ for an anisotropic three-dimensional superconductor. Our results thus suggest that at these low temperatures, the superconductivity in our samples is in the two-dimensional limit.

Examining Fig. 1**c** data at higher magnetic fields, we find that the magnetoresistivity traces exhibit distinct $1/\mu_0 H$-periodic quantum oscillations (*i.e.*, the Shubnikov de Haas-effect), which are particularly prominent at small $\theta$ angles. Through careful analysis of the quantum oscillatory data (provided in Extended Fig. 2, with a detailed discussion in Methods Section V), we observe that the frequencies of the quantum oscillations follow a $1/\cos\theta$ dependence, characteristic of quasi-two-dimensional Fermi surfaces. Therefore, the anisotropic character of the superconducting state of CsV$_3$Sb$_5$ reflects the anisotropy of its electronic structure.

The angular-dependent measurements provide compelling evidence for the two-dimensional nature of the superconductivity in CsV$_3$Sb$_5$. Remarkably, we also observe a large in-plane upper critical field ($\mu_0 H_{c2,||}$) of almost 10 T. In conventional superconductors, the presence of a strong external magnetic field suppresses superconductivity through either the orbital[26] or (spin) Zeeman effect[27,28]. While in most cases the orbital effect dominates, it becomes less dominant for magnetic fields applied parallel to the planes of two-dimensional superconductors. If the orbital effect became negligible, the upper limit on $\mu_0 H_{c2}$ would be primarily imposed by the Zeeman effect. This upper limit, known as the Pauli (or Clogston–Chandrasekhar) paramagnetic limit ($\mu_0 H_p =$ 1.84 $\times T_c$ T/K for weakly coupled BCS superconductors)[27,28], is 6.44 T in our case (marked in Fig. 1**d**). Note that if CsV$_3$Sb$_5$ was indeed Pauli limited for in-plane fields, we would expect a deviation from the Tinkham formula. Given the good fit in Fig. 1**d**, we conclude that the critical field of CsV$_3$Sb$_5$ exceeds the Pauli limit by at least 50%, exceeding 9.8 T. A high $\mu_0 H_{c2,\parallel}$ value is often associated to either unconventional superconductivity, for example, emerging from strong electronic correlations, or disorder leading to strong spin-orbit scattering, which is expected to provide a modest increase in $\mu_0 H_{c2,\parallel}$. The violation of the Pauli limit suggests that CsV$_3$Sb$_5$ is either in the dirty-limit, or the superconducting state is unconventional in nature, as was discussed in the context of few-layer transition metal dichalcogenides[29-32] and twisted bilayer graphene[33] which are suspected to exhibit unconventional superconductivity.

After investigating the anisotropy of $H_{c2}$ from the *c*-axis to the *ab*-plane, our attention turns to examining the evolution of $H_{c2}$ within the *ab*-plane as we vary the angle $\varphi$ (illustrated in Fig. 1**e**). By tuning $\varphi$, we observe a progressive shift in the superconducting transition along the magnetic field axis, resulting in anisotropic $H_{c2,\parallel}$ values (Fig. 1**e**). To eliminate the possibility of an accidental out-of-plane field component, we conducted careful angular-dependent measurements, as detailed in Methods Section VI and Extended Fig. 3. In this analysis, we define $\varphi = 0°$ as the in-plane angle at which $H_{c2,\parallel}$ reaches its maximum value. It should be noted that $\varphi = 0°$ does not align with the direction of the electrical current, but with the minimum of the angular magnetoresistivity (maximum of $H_{c2}$). Summarizing the in-plane rotation experiments, Fig. 1**f** depicts $H_{c2}$ as a function of $\varphi$ in a polar plot. Remarkably, we find that $H_{c2}(\varphi)$ exhibits an emergent two-fold symmetry as a function of $\varphi$; see Methods Section VII for details on the observed anisotropy. This finding is intriguing from a crystal structure perspective since the underlying lattice possesses a six-fold rotational symmetry. However, as discussed earlier, the normal state already displays an emergent two-fold symmetric behavior (see Extended Fig. 4), which appears to be inherited by the superconducting state, giving rise to a two-fold symmetric response[15].

To gain further insights into the phase diagram for different orientations of the magnetic field, we conducted systematic temperature-dependent measurements. Figures 2**a**,**b** and 2**d**,**e** display the obtained transport data with the magnetic field applied parallel and perpendicularly to the sample plane, respectively. In both cases, the superconducting transition gradually shifts to lower magnetic fields as the temperature is increased. The temperature dependence of $H_{c2,\parallel}$ and $H_{c2,\perp}$ are summarized in Figs. 2**c** and 2**f**, respectively (Refer to Extended Fig. 5 for the $\mu_0 H_{c2}$ as a function of *T* phase diagrams using different criteria to define $\mu_0 H_{c2}$. These diagrams consistently show the same features as in Figs. 2**c** and 2**f**, regardless of the $\mu_0 H_{c2}$ definition used.) Close to the critical temperature, the temperature dependence of the upper critical field is well described within the Ginzburg-Landau description of two-dimensional superconductivity (see Methods Section IV). We find a zero-temperature in-plane coherence length $\xi_{\text{GL}}(0) \simeq 17$ nm, which is in good agreement with previously reported values[34] and a superconducting thickness $d_{SC} \approx 8$ nm. Additionally, we obtain $\xi_{\text{GL}}(0) \simeq 13.6$ nm and $d_{SC} \simeq 8.6$ nm directly from $H_{c2,\perp}(T = 0)$ and $H_{c2,\parallel}(T = 0)$, respectively. However, these values, which assume a purely orbitally limited superconductor at $T = 0$, may only serve as upper bounds due to the Pauli limit violation observed by us. Notably, $\xi_{\text{GL}}(0)$ is about 25 to 30 times larger than the in-plane lattice constant, $a = 5.5$ Å, and larger than $d_{SC}$. Despite the good agreement with the fit to two-dimensional superconductivity, it should be noted that the extracted superconducting thickness is much smaller than the sample thickness but nearly 10 times larger than the interlayer spacing $t = 0.93$ nm of CsV$_3$Sb$_5$.

The most striking feature of the phase diagram, however, is a pronounced upturn of the upper critical field around 1 K. Similar upturns have been observed in several systems, where it is associated with a field-induced phase[35,36] or a dimensional crossover from three to two dimensions[37]. However, we observe an upturn irrespective of the field orientation, which is incompatible with either of these scenarios. We argue, supported by our heat capacity and thermal conductivity measurements, that a possible explanation for our phase diagram is the opening of a second (nearly) decoupled gap on a previously non-superconducting band in the low-temperature regime with a much higher critical field. This high critical field in the low-temperature regime (Figs. 2**c** and 2**f**) could be due to a much shorter coherence length, or the possibility of the smaller gap being in the dirty limit, akin to the situation already discussed for $MgB_2$[38,39]. Consequently, we expect substantial residual degrees of freedom in the superconducting regime observed at higher temperatures.

We further substantiate our electrical transport measurements by performing systematic thermal transport measurements, which are well-suited for studying bulk superconductivity. Unlike electrical resistivity, thermal conductivity may remain non-zero in the superconducting state due to delocalized quasiparticles because Cooper pairs, which do not carry entropy, do not contribute to thermal transport[40]. Consequently, the thermal conductivity allows us to examine delocalized low-energy quasiparticle excitations and their response to magnetic fields[40]. We employed the conventional one-heater three-thermometer method on an exfoliated $CsV_3Sb_5$ sample (see Methods Section VIII for details). Figure 3**a** demonstrates that $\kappa_{xx}/T$ exhibits a distinct transition around $T \simeq 3$ K, which is near the observed $T_c$ according to our electrical transport experiments (Fig. 1**b**). As the temperature is reduced well below $T_c$, $\kappa_{xx}/T$ saturates before a sharp change in the slope occurs around 0.8 K (Fig. 3**a** inset), or near the boundary between the low-temperature and high-temperature regimes according to our electrical transport experiments (Fig. 2). Notice the near constant value of $\kappa_{xx}/T$ deep inside the superconducting regime I (Fig. 3**a** inset), implying either (*i*) the presence of residual quasiparticles due to an ungapped Fermi surface, or (*ii*) a gap with nodal structure. In any case, this change in slope is consistent with a scenario based on two-gap superconductivity with nearly decoupled gaps. Further data and discussion on the thermal conductivity, including its field dependence, are presented in Extended Fig. 6 and Methods Section IX.

Both anomalies observed in the thermal conductivity, at the superconducting transition and at the boundary between both superconducting regimes, are replicated by heat-capacity measurements at zero field as a function of the temperature (Fig. 3**c**); see Methods Section X for details. Figure 3**c** contains two datasets, corresponding to the heat capacity normalized by the temperature collected at zero-field, or $C/T_{\mu_0 H= 0\ T}$, but subjected to two different background or lattice contribution subtractions, in order to evaluate solely the electronic contribution to the heat capacity. To subtract the background, two measurements were performed, $C/T$ as a function of $T$ under $\mu_0 H = 2$ T || $c$-axis and $\mu_0 H = 9$ T || $ab$-plane of the crystal, with both values being above and very close to $\mu_0 H_{c2,\perp}$ and $\mu_0 H_{c2,\parallel}^b$, respectively. Both subtractions yield two anomalies. The middle point of the first anomaly is located at $T \cong 3$ K, with the associated peak occurring at a lower temperature of $\cong 2.66$ K. Therefore, it corresponds to the anomaly associated with the onset of superconductivity. A second anomaly is seen just below $T = 1$ K, showing a very modest peak around $T = 0.6$ K. The extrapolation of the heat capacity to $T = 0$ K suggests the near absence of residual electronic contribution. This finding will have to be confirmed by extending the heat capacity measurements to temperatures below our limit of $T \approx 200$ mK. The change in the heat capacity across the initial superconducting transition $\Delta C$ normalized by the electronic contribution measured above the transition, $\gamma_e \cong 27$ mJ/molK$^2$, yields $\Delta C/\gamma_e \cong 0.6$, which is significantly reduced when compared to the BCS value $\Delta C/\gamma_e \cong 1.42$. This reduction indicates that a large fraction of the electronic states remains in the normal state at the primary transition, again consistent with our scenario of nearly decoupled gaps.

To corroborate the two-gap scenario, we calculated both the longitudinal heat transport and specific heat in a minimal model of $CsV_3Sb_5$ with two weakly coupled superconducting gaps on different Fermi surfaces, see Fig. 3**b** and **d**, respectively. In particular, we consider the circular pocket around the Γ-point, which derives mostly from Sb $p$ orbitals and the hexagonal pocket, mostly stemming from V $d$ orbitals[7]. Furthermore, we assume a primary gap on the Sb-derived band with a secondary gap on the V-derived band[23]. For simplicity, we neglect the band folding due to the charge density wave. For nearly decoupled superconducting gaps on these two bands, we indeed observe a temperature dependence consistent with the measurements (see SI for details of these calculations).

Finally, we explore the angular dependence of the planar thermal conductivity, specifically $\kappa_{xx}/T$ for fields rotating within the conducting planes, as $CsV_3Sb_5$ transitions from the superconducting to the metallic state with increasing magnetic fields. In general, we expect three possible sources of anisotropy for $\kappa_{xx}/T$: (1) at fixed field, the anisotropy of $H_{c2}$ results in a field-direction-dependent critical temperature. Combined with a strong temperature dependence of $\kappa_{xx}/T$, an anisotropy with a maximum in $\kappa_{xx}/T$ along the direction of the highest $H_{c2}$ is expected. (2) In the mixed state, a two-fold symmetric signal is expected with a maximum for fields along the transport direction. (3) $\kappa_{xx}/T$ is finally angle sensitive for (near) nodal superconducting gap structures through the density of heat-carrying quasiparticles and the Doppler shift of quasiparticle excitations outside of vortex cores, the so-called Volovik effect[41,42] (see Methods Section XI for details). Therefore, by measuring the angular dependence of $\kappa_{xx}/T$ one can evaluate the anisotropy of the superconducting gap function or even detect the presence of nodes, as done for the cuprate superconductors[43], $UPt_3$[44], or $Sr_2RuO_4$[40,45,46]. In Fig. 4, we provide $\kappa_{xx}/T$ as a function of the planar angle $\phi$ between the external magnetic field and the direction of the thermal gradient $\nabla T$, revealing a modest anisotropy in $\kappa_{xx}(\phi)/T$, in other words comparable to the anisotropy between the maximum and minimum values of the in-plane upper critical field $\mu_0 H_{c2,||}(\phi)$ (Fig. 1**e**). This is consistent with a superconducting gap that is mildly anisotropic. While we cannot determine the direction of the maximum critical field, it seems plausible that the two directions align in the high-temperature regime. Intriguingly, however, the thermal-transport anisotropy rotates by 90° when the temperature is decreased and the superconducting state transitions from the high-temperature to the low-temperature regime. Furthermore, as the direction of the upper-critical-field anisotropy does not rotate between the two regimes (see Extended Fig. 7 and Methods Section XII), sources (1) and (2) cannot explain the observed anisotropy. As such, we must assume that the anisotropy is caused by a strong gap anisotropy of at least one of the gaps, potentially even with nodes.

Nodes in one of the gaps would be consistent with the observation of a residual density of states in scanning tunneling spectroscopy[13,14], the coexistence with a time-reversal-symmetry-broken charge density wave state[47], and the lack of inversion or in-plane mirror symmetry[48]. This would also be consistent with previous claims of nodal superconductivity in the sister compounds $RbV_3Sb_5$ and $KV_3Sb_5$ via muon scattering[18]. However, it is worth noting that in ref.[49], an anisotropic $s$-wave gap has been suggested for low-impurity $CsV_3Sb_5$. While we cannot discard the possibility of a heavily anisotropic gap, as presented in Section II of the SI, Figs. 3**a** and 3**c** cannot be reproduced by a heavily anisotropic $s$-wave function for the main gap alone.

In summary, by employing a combination of electrical transport, thermal transport, and heat capacity measurements, we unveiled two superconducting regimes in $CsV_3Sb_5$, which we associate to nearly decoupled two-gap superconductivity. A substantial residual thermal conductivity is observed deep in the superconducting regime observed at higher temperatures, likely resulting from ungapped Fermi surface sheets[14], although we cannot discard that it could correspond to residual quasiparticle excitations due to a nodal gap structure. We also observe a 90º rotation in the anisotropy of the planar thermal conductivity measured under a rotating magnetic field, when cooling from the higher-temperature superconducting regime towards the lower-temperature one. As the direction of the upper-critical-field anisotropy does not simultaneously rotate between both regimes, we propose that the anisotropy in the thermal conductivity at low temperatures is induced by a strong anisotropy inherent to at least one of the gaps, noting that it could also be a manifestation of nodal gap structure. Nodes would be consistent with the residual density of states observed in several scanning tunneling experiments, but difficult to reconcile with other studies such as temperature-dependent penetration depth measurements Clearly, more complete studies of, for example, thermal conductivity at very low temperatures along with a comprehensive modeling of the electron-phonon scattering within the superconducting state, will be required to fully understand the nature of the superconducting pairing symmetry in $CsV_3Sb_5$. Our results show that a band-selective pairing with two nearly decoupled gaps are the scenario on which further analysis needs to be based.

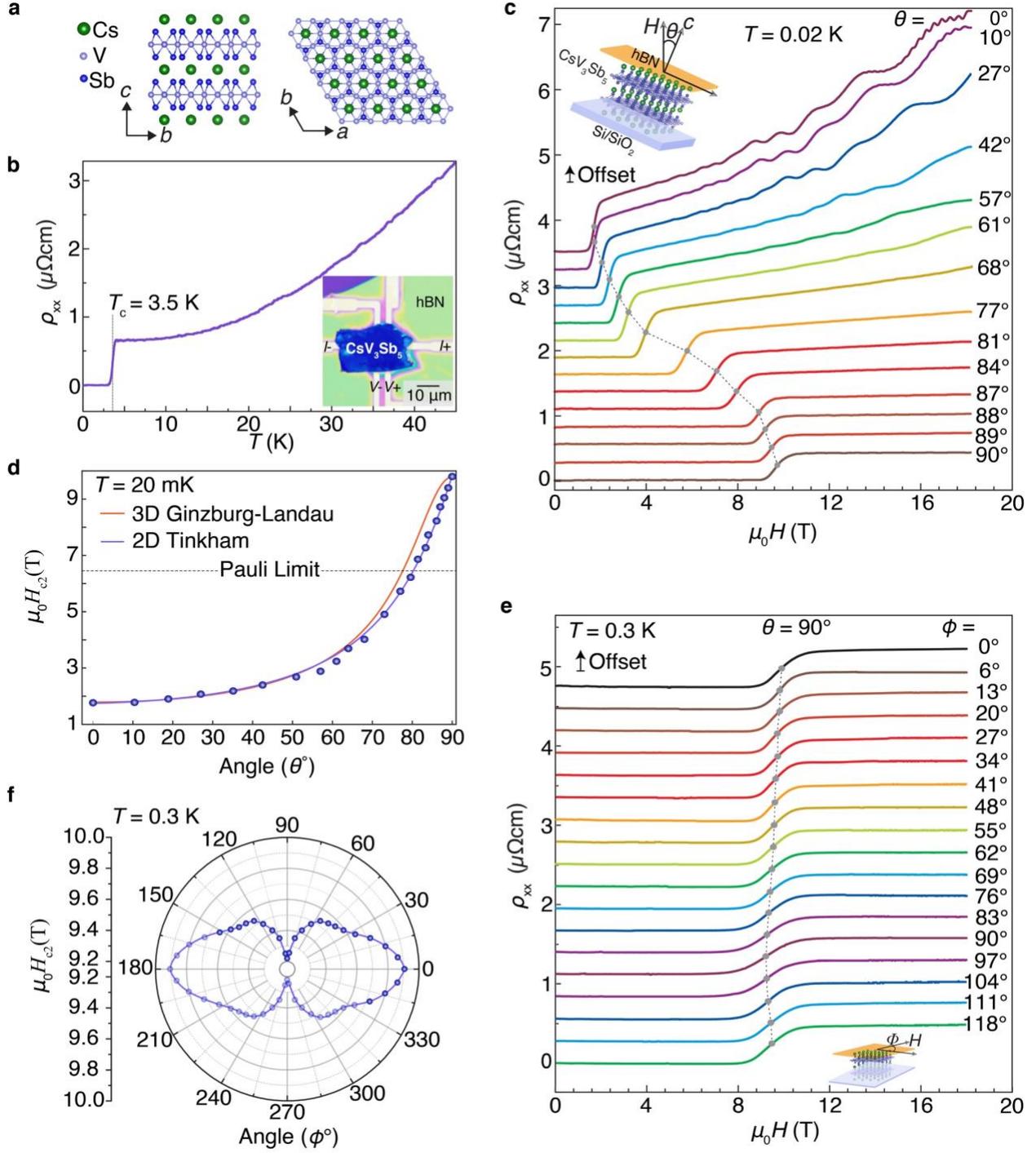

**Fig. 1: Observation of two-dimensional and two-fold symmetric superconductivity in $CsV_3Sb_5$ visualized through its upper critical field. a,** Crystal structure of $CsV_3Sb_5$, featuring a layered, quasi two-dimensional crystal structure along the *c*-axis and forming a six-fold symmetric V kagome lattice within the V-Sb slab. **b,** Four-probe resistivity of a mechanically exfoliated ∼90 nm thick $CsV_3Sb_5$ crystal, exhibiting metallic behavior as a function of the temperature (*T*) before transitioning to a superconducting state with a $T_c$ of 3.5 K. Inset shows an optical microscopy image of the device comprising a $CsV_3Sb_5$ flake encapsulated with a *h*-BN flake, indicating current and voltage leads. **c,** Resistivity as a function of the magnetic field ($\mu_0H$) at $T \simeq 0.02$ K for different out-of-plane

rotation angles ($\theta$). The inset illustrates the direction of rotation, where $\theta$ represents the angle between the magnetic field and the $c$-axis of the crystal. For each trace, the upper critical magnetic field ($\mu_0 H_{c2}$), defined as the field at which the resistivity becomes half of its value in the normal state, is denoted by grey circles. Traces corresponding to different $\theta$ angles are vertically offset for improved visibility. **d**, $\theta$-dependence of the upper critical field. Purple and red curves represent fits to the data using the Tinkham formula for a two-dimensional superconductor and the three-dimensional Ginzburg-Landau anisotropic mass model, respectively. The data exhibits excellent agreement with two-dimensional superconductivity. The grey dashed line indicates the Pauli paramagnetic limit. **e**, Resistivity as function of the magnetic field at $T \simeq 0.3$ K for different in-plane rotation angles ($\varphi$). The inset illustrates the plane of rotation, where $\varphi$ denotes the angle between an in-plane magnetic field along the $ab$ plane ($\mu_0 H_{c2,\parallel}$) and the axis leading to the largest value of $\mu_0 H_{c2,\parallel}$. For each trace, the value of $\mu_0 H_{c2,\parallel}$ is indicated by grey circles. Traces corresponding to different $\varphi$ angles are vertically offset for clarity. **f**, $\varphi$-dependence of $H_{c2,\parallel}$, revealing an emergent two-fold symmetry within $ab$-plane despite the six-fold one of the underlying lattice. The darker (lighter) color symbols denote raw (repeated assuming 180° periodicity) data. We used normal state $\rho_{xx}$ values over a 1 T field range just above the transition to determine the normal state resistivity ($R_{\text{Normal}}$) and the corresponding range of $H_{c2}$ values obtained from $\rho_{H_{c2}} = 0.5 \times \rho_{\text{Normal}}$ for error bars. For panel **d** (**f**), the error bar is ~±0.03 T (~±0.01 T), smaller than the symbol size.

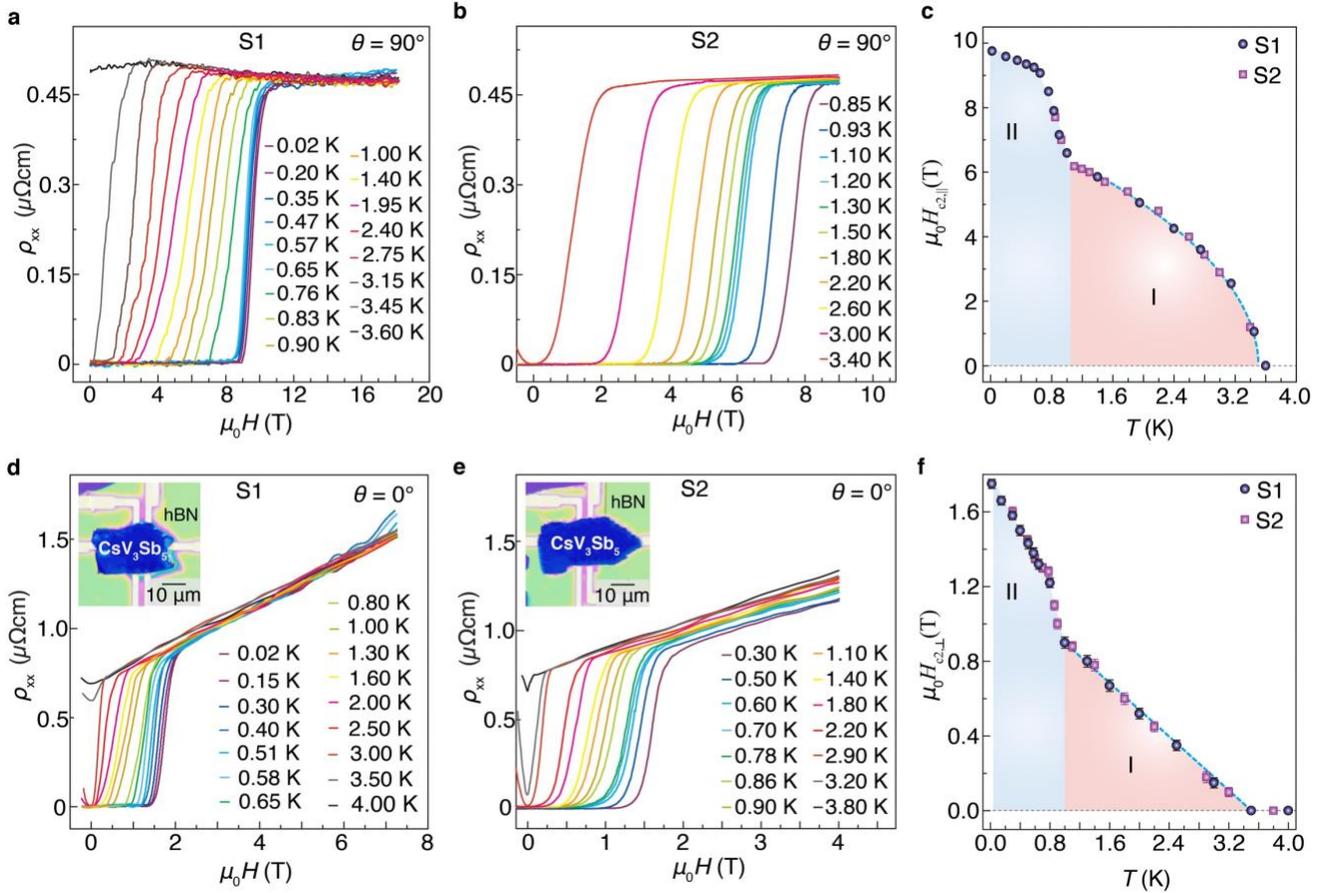

**Fig. 2: Evidence for two distinct superconducting regimes. a**, **b**, Resistivity $\rho_{xx}$ as a function of an in-plane magnetic field $\mu_0 H_\parallel$ ($\theta = 90°$) for several temperatures. The data was obtained from two samples, S1 (panel **a**) and S2 (panel **b**) having similar thicknesses. **c**, In-plane upper critical field $\mu_0 H_{c2,\parallel}$, extracted from the traces in panels **a** and **b**, plotted as a function of the temperature $T$. Two clear regimes (I and II) comprising two distinct temperature regions are observed. The data acquired from the two samples reveals an overlapping phase-diagram. Blue dashed curve represents a fit to the Ginzburg-Landau model for two-dimensional superconductors corresponding to the high temperature regime near $T_c$. **d, e**, Resistivity as a function of an out-of-plane magnetic field $\mu_0 H_\perp$ ($\theta = 0°$) for different temperatures, obtained from samples S1 (panel **d**) and S2 (panel **e**). Optical microscopy images of samples S1 and S2 are shown in the insets of panels **d** (same device as in the inset of Fig. 1**b**) and **e**, respectively. **f**, Out-of-plane upper critical field $\mu_0 H_{c2,\perp}$ as a function of $T$ as extracted from the traces in panels **d** and **e**. This phase diagram also features two distinct regimes (I and II) displayed in panel **c**. The data is obtained from samples S1 and S2. Blue dashed curve represents a fit of the high temperature regime to the Ginzburg-Landau model for two-dimensional superconductors (with $T_c = 3.5$ K as obtained in Fig. 1**b**). $\mu_0 H_{c2}$ is defined using $\rho_{H_{c2}} = 0.5 \times \rho_{\text{Normal}}$, where $\rho_{\text{Normal}}$ is the normal state resistivity just above the superconducting transition. To extract $\rho_{\text{Normal}}$, we fitted $\rho_{xx}$ to a straight line, subsequently choosing its deviation from linearity as the value of $\rho_{\text{Normal}}$. The error bar is given by the uncertainty on the exact field value where the deviation from linearity occurs. For $\mu_0 H_{c2,\parallel}$, the error bar is ~±0.01 T, smaller than the symbol size in panel **c**. For $\mu_0 H_{c2,\perp}$, the error bar is ~±0.03 T, matching the symbol size in panel **f**. Hence, we included error bars in panel **f**.

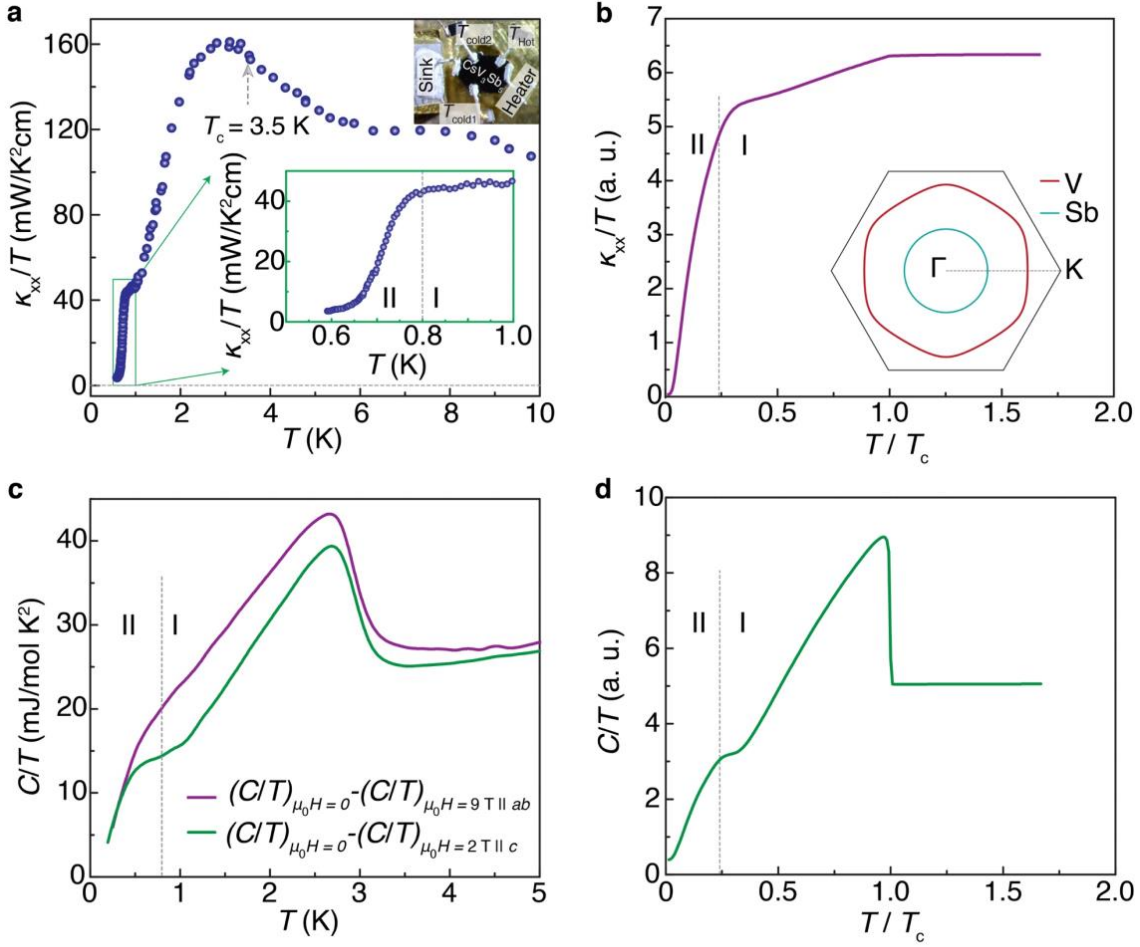

**Fig. 3: Thermodynamic and thermal transport evidence for two superconducting regimes. a,** Thermal conductivity ($\kappa_{xx}/T$) under zero magnetic field across the superconducting transition as a function of the temperature for thermal gradients applied within the *ab*-plane. Gray vertical arrow indicates the $T_c$ obtained from the electrical transport measurements. Top inset: picture of the experimental setup, depicting the thermal contacts made on a single crystal of $CsV_3Sb_5$ indicating heater, heat sink, and thermometers. Bottom inset: $\kappa_{xx}/T$ as a function of $T$ for $T \ll T_c$. An anomaly is observed near $T \simeq 0.8$ K, which likely corresponds to the boundary between regimes I and II. The location in temperature of this anomaly is consistent with the phase diagrams based on our electrical transport measurements (Figs. 2 and 3). **b,** A two-gap model on distinct Fermi surface sheets captures the overall behavior displayed by $\kappa_{xx}/T$ (see SI). Note that we only calculated electronic contributions, which do not account for the large decrease of the thermal transport below $T_c$. This decrease could possibly stem from phononic contributions. **c,** Electronic contribution to the heat capacity across the superconducting transition. To obtain the electronic contribution to $C/T$, the heat capacity as a function of the temperature was measured above and near $\mu_0 H_{c2}^c$ and $\mu_0 H_{c2}^{ab}$ respectively, or under $\mu_0 H^c = 2$ T and $\mu_0 H^{ab} = 9$ T, and subsequently subtracted from the $\mu_0 H = 0$ T trace. Two anomalies are observed, coinciding with the locations in temperature of the I and II superconducting regimes. We refer to Methods Section X and Extended Fig. 8 for details. **d,** Specific heat $C/T$ calculated using a two-gap model, exhibiting the peak feature at $T_c$ and a secondary feature when the nearly decoupled gap opens (See SI).

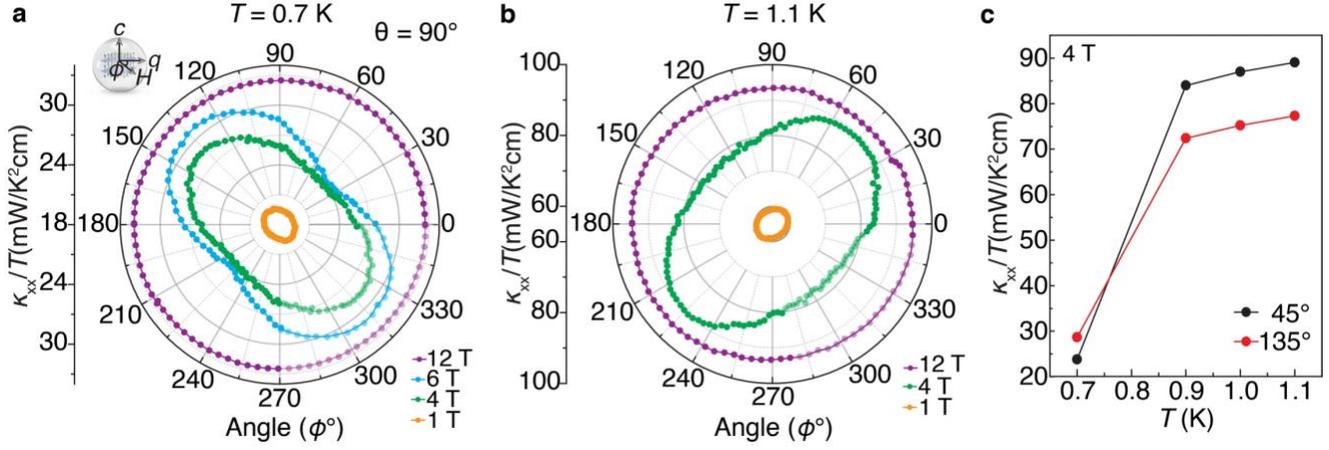

**Fig. 4: Angular dependence of the in-plane thermal conductivity within the I and II regimes. a**, Polar plot of the in-plane thermal conductivity under fixed magnetic fields of $\mu_0 H = 1, 4, 6,$ and 12 T, as the field is rotated within the *ab*-plane as a function of the planar angle $\phi$ at $T \simeq 0.7$ K (regime II). The angle $\phi$ represents the orientation between the in-plane magnetic field and the direction of the applied heat current ($q$), as shown in the inset. The data displays two-fold symmetric oscillations, that are particularly prominent at higher magnetic fields (for fields below $H_{c2,||}$). For fields above $H_{c2,||}$, such as at 12 T, the anisotropy becomes negligible. **b**, Polar plot of the *ab*-plane thermal conductivity at fixed magnetic fields of $\mu_0 H = 1, 4,$ and 12 T, as the field is rotated within the *ab*-plane as a function of $\varphi$ at $T \simeq 1.1$ K (regime I). As for the panel **a** data, clear two-fold symmetric oscillations are observed for fields below $H_{c2,||}$. At $\mu_0 H =12$ T, which is above $H_{c2,||}$, no visible anisotropy is observed. A clear $\sim \pi$ phase shift is observed between the 0.7 K and 1.1 K data as is evident from the data in panels **a** and **b**. The darker (lighter) color symbols in panels **a** and **b** denote raw (repeated assuming 180° periodicity) data. We refer to Methods Section XII and Extended Fig. 9 for details. **c**, Amplitude of the *ab*-plane thermal conductivity under a fixed magnetic field of $\mu_0 H = 4$ T for two angles $\phi = 45^0$ (black markers) and $135^0$ (red markers) as a function of the temperature. The position in $\phi$ where the maximum is observed in the thermal conductivity switches from $\phi = 45^0$ to $135^0$ when cooling down to $T = 0.7$ K from 0.9 K.

## Methods:

### I. Synthesis of CsV$_3$Sb$_5$ single crystals

Single crystals of CsV3Sb5 were grown using a flux method. Cs (99.8% purity) liquid, V (99.999% purity) bulk, and Sb (99.9999% purity) shots were used as starting materials. A flux of Cs0.4Sb0.6 was prepared by mixing the components. The mixture was then placed in an alumina crucible and sealed in a quartz ampoule within an Argon-filled glovebox. It was heated to 1000 °C over a period of 200 hours and soaked for 24 hours at that temperature. Subsequently, the mixture was slowly cooled at a rate of 3.5 °C/h. After turning off the muffle oven, the crystals were immersed in deionized water to remove the flux. This process yielded shiny, hexagonal single crystals of CsV$_3$Sb$_5$. X-ray diffraction on these samples revealed a Bragg peak with a width of just $\simeq 0.07$ degrees[50]. The transition near $T = 90$ K in resistivity is remarkably sharp, as evidenced by the peak in its derivative, with a similar anomaly observed in magnetization. These results indicate that the material is clean and well-ordered. Furthermore, our samples are of sufficiently high quality to exhibit quantum oscillatory phenomena at relatively low magnetic fields. We found that samples from any given batch are similar in quality. Disorder should average both gaps, making it difficult to expose their anisotropy. Hence, the anisotropy observed in the thermal conductivity, by itself, would demonstrate that our crystals are rather clean. However, the extent to which the anisotropy reported here persists in the presence of impurities once two superconducting gaps open remains as an open question.

## II. Device fabrication

We utilized a polydimethylsiloxane (PDMS) stamp-based mechanical exfoliation technique to fabricate $CsV_3Sb_5$ samples. We patterned the electrical contacts onto the sample exfoliated onto the silicon substrates with a 280 nm layer of thermal oxide using electron beam lithography, followed by chemical development and metal deposition (5 nm Cr/35 nm Au). The fresh $CsV_3Sb_5$ samples were mechanically exfoliated from bulk single crystals (on the crystallographic *ab* plane) on PDMS stamps. It is worth noting that we do not know the in-plane crystallographic orientations of the samples, *i.e.*, or the directions of the *a*- and *b*- axes. Prior to transferring the $CsV_3Sb_5$ flakes onto the $SiO_2$/Si substrates with pre-patterned Cr/Au electrodes, we visually inspected the crystals under optical microscopy to select samples with favorable geometries. The approximate thickness of the flakes was determined *via* optical contrast, a commonly used method for sensitive samples[51]. This approach relies on the fact that samples with different thicknesses exhibit distinct optical contrasts[52]. Initially, we characterized the thickness of the samples through atomic force microscopy and established the corresponding relationship between the optical contrast and the thickness. To preserve the intrinsic properties of the compound and minimize environmental effects, we encapsulated the samples with exfoliated *h*-BN layers having thicknesses ranging from approximately 10 to 30 nm, which ensured that the samples remained protected from direct air exposure. The entire sample fabrication process was conducted within a glovebox equipped with a gas purification system, maintaining an environment with extremely low levels of oxygen and water vapor (<1 ppm).

## III. Electrical transport measurements

Electrical transport measurements requiring high magnetic fields were conducted using two separate systems: a dilution refrigerator coupled to a magnet capable of generating magnetic fields up to 18 T, and a $^3$He system equipped with a Bitter magnet capable of reaching magnetic fields up to 41 T. These measurements were performed at the National High Magnetic Field Laboratory in Tallahassee, Florida, USA. Note that, in our measurements shown in Figs. 1 and 2, we used a field sweep rate of 0.2 T/min and a time constant of 1 s, resulting in a 3.3 mT interval between neighboring data points. This interval is smaller than the symbol size used in the plots showing discrete data points. Furthermore, the temperature uncertainty is below 1%, which is also smaller than the symbol size. For the electrical transport measurements shown in Extended Fig. 3, a Physical Property Measurement System was used, providing temperature control down to 1.8 K and magnetic fields up to 9 T. To ensure precise alignment of the samples with respect to the external magnetic field, the devices were mounted on a rotator that allowed for in situ sample rotation. Multiple devices were prepared and measured during multiple measurement runs to ensure the reproducibility of the results.

## IV. $H_{c2}$ and two-dimensional superconductivity

In this section, we delineate the formulas that we used to fit the angular and temperature dependences of the upper critical magnetic fields for fields perpendicular and parallel to the sample plane.

To fit the angular dependence results, we used the following two formulas:

1. Tinkham formula: The Tinkham formula[25] is given by $\left|\frac{H_{c2}(\theta)\cos\theta}{H_{c2,\perp}}\right| + \left(\frac{H_{c2}(\theta)\sin\theta}{H_{c2,\parallel}}\right)^2 = 1$. Here, $H_{c2,\perp}$ and $H_{c2,\parallel}$ are the upper critical magnetic fields for fields perpendicular and parallel to the sample plane, respectively. $\theta$ denotes the angle between the magnetic field direction and the out-of-plane orientation.

2. Ginzburg-Landau anisotropic mass model: Ginzburg-Landau anisotropic mass model[25] describes the angular dependence of $H_{c2}$ for an anisotropic three-dimensional superconductor as: $\left(\frac{H_{c2}(\theta)\cos\theta}{H_{c2,\perp}}\right)^2 + \left(\frac{H_{c2}(\theta)\sin\theta}{H_{c2,\parallel}}\right)^2 = 1$.

To fit the temperature dependence results, we used the Ginzburg-Landau description of two-dimensional superconductivity[25]:

- $H_{c2,\parallel}(T) = \frac{\Phi_0\sqrt{3}}{\pi\xi_{GL}(0)d_{sc}}\left(1 - \frac{T}{T_c}\right)^{1/2}$ for in-plane fields,
- $H_{c2,\perp}(T) = \frac{\Phi_0}{2\pi\xi_{GL}(0)^2}\left(1 - \frac{T}{T_c}\right)$, for out-of-plane fields, which displays a linear dependence on temperature.

Here $\Phi_0$ denotes the magnetic flux quantum and $T_c$ is the critical temperature.

It is important to note that the Ginzburg-Landau description is only valid near $T_c$. Therefore, we applied it exclusively to the high-temperature regime (Regime I). The system parameters for the low-temperature regime cannot be extracted from the Ginzburg-Landau fits.

Finally, we note that multiband superconductors can exhibit both a linear or square-root temperature dependence of $\mu_0 H_{c2}(T)$ depending on the relative importance of orbital and paramagnetic limiting[53]. Therefore, the linear $\mu_0 H_{c2,\perp}(T)$ observed in our experiments (Fig. 2**f**) may indicate multiband superconductivity. However, we believe multiband physics is unlikely to be the leading factor. The reason is that (1) within a multiband picture, a second feature only appears for out-of-plane fields, as the in-plane fields need to be paramagnetically limiting for the square-root behavior and (2) the secondary feature at $T \sim 0.8$ K is (nearly) independent of magnetic-field direction, which is unexpected for a multiband scenario, where the anisotropies of the two bands are independent. Hence, we did not invoke multiband physics to fit the $\mu_0 H_{c2}(T)$ data at higher temperatures. In fact, our data points to two nearly decoupled superconducting gaps, implying that the higher temperature portion of the phase diagram is dominated by a single gap (see Supplementary Information, Sec. III).

## V. Quantum Oscillations in CsV$_3$Sb$_5$

Upon careful examination of the angular-dependent magnetoresistivity traces in Fig. 1**c**, we observe clear quantum oscillations, characterized by $1/\mu_0 H$-periodicity in the magnetoresistivity. In this section, we investigate how these quantum oscillations evolve as we vary the magnetic field orientation from the crystallographic *c*-axis to the *ab*-plane. To analyze the quantum oscillations, we first extract the oscillatory component of the magnetoresistivity, $\Delta\rho_{xx}$, by subtracting a smooth background that was fit to a polynomial. The resulting data is plotted as a function of $1/\mu_0 H$ in Extended Fig. 2**a**, illustrating the $1/\mu_0 H$-periodic quantum oscillations at various angles ($\theta$) between the magnetic field and the *c*-axis. These traces exhibit well-defined oscillations for $\theta$ values ranging from 0° ($\mu_0 H \parallel c$) to 68°. Next, we perform a Fourier transform of the $\Delta\rho_{xx}$ data, yielding a series of Fourier transform spectra as a function of $\theta$, presented in Extended Fig. 2**b**. For each $\theta$ angle when $\theta \leq 68°$, we observe two distinct peaks associated to the $1/\mu_0 H$ periodicity. These peaks, marked with color-coded arrows in Extended Fig. 2**b**, are associated to Fermi surface cross-sectional areas, according to the Onsager relation. The observed peak frequencies exhibit a reasonable agreement with the frequencies reported in previous studies[54]. By tracking the angular dependence of the frequency associated with both peaks or Fermi surface cross-sectional areas, we extract their angular dependence, as shown in Extended Fig. 2**c**. Remarkably, the evolution of both Fourier-transform peak frequencies as a function of $\theta$ follows a $1/\cos\theta$ behavior, indicating the two-dimensional character of the Fermi surfaces of CsV$_3$Sb$_5$. It is important to note that for $\theta$ values above 68°, we do not observe clear quantum oscillations. This is a direct consequence of the two-dimensionality of the Fermi surfaces in CsV$_3$Sb$_5$. In our limited magnetic field range, for angles $\theta > 68°$, the perpendicular magnetic field ($\mu_0 H \parallel c$) component, which is responsible for inducing quantum oscillations in the two-dimensional Fermi surfaces lying along the *ab* plane, becomes very small and is insufficient to generate well-developed quantum oscillations.

## VI. In-plane rotation measurements

To rule out any potential influence of an accidental out-of-plane component of the magnetic field, we conducted continuous in-plane rotation measurements on the same sample with and without an intentional 6-degree canting. Specifically, we performed measurements at $\theta = 90°$ (without canting) and at $\theta = 84°$ (with canting). The angular ($\varphi$) variation of the magnetoresistivity traces at different magnetic fields for two temperatures is shown in Extended Fig. 3. In line with the convention in Figs. 1**e** and 1**f**, $\varphi$ represents the angle between the in-plane magnetic field along the *ab* plane ($\mu_0 H_{c2,\parallel}$) and the axis with the largest $H_{c2,\parallel}$. As depicted in Extended Figs. 3**a** and 3**b**, we observe a consistent two-fold modulation in the magnetoresistivity across all the magnetic fields shown. The modulation is particularly prominent within the field range encompassing the onset and offset of superconductivity. For example, at $T = 1.8$ K (Extended Fig. 3**a**), the modulation strength is most significant between 4-6 T magnetic fields, which are close to $\mu_0 H_{c2}$ (around 5.5 T) and correspond to the transition region between the zero-resistivity state and the normal state. Although attenuated in the normal state, the modulation remains present. This behavior is consistent with the reported two-fold symmetry of the nematic charge order observed in the normal state. A qualitatively similar trend is observed at $T = 2.8$ K (Extended Fig. 3**b**), with the strongest modulation occurring for fields between 2.5-5 T, again near $\mu_0 H_{c2}$ (around 3.4 T), encompassing the transition region between the zero-resistivity and normal states, and maintaining a two-fold modulation in the normal state albeit with reduced strength. The two-fold modulation can be accurately described by a sinusoidal function of the form $sin[2(\varphi + \varphi_0)]$, where $\varphi_0$ represents the relative phase from $\varphi = 0°$. This functional form is evident in the fitting curves plotted using dashed lines in Extended Figs. 3**a** and 3**b**. The observed two-fold symmetry is also consistent with the data presented in Figs. 1**e** and 1**f**. Specifically, the trace corresponding to $\varphi = 0°$ exhibits the highest $H_{c2}$, as corroborated by the magnetoresistivity minimum, whereas the trace corresponding to $\phi = 90°$ displays the lowest $\mu_0 H_{c2}$, as corroborated by the magnetoresistivity maximum.

In the case of intentional canting, achieved by mounting the sample on a wedge (as depicted in the inset of Extended Fig. 3**c**), the two-fold modulation is still present, but the magnetoresistivity traces exhibit noticeably different shapes that can be well fit by $|sin(\varphi + \varphi_0)|$ (where $\varphi_0$ represents the relative phase from $\varphi = 0°$), as indicated by the dashed fitting curves in Extended Figs. 3**c** and 3**d**. This distinct shape of the curves resulting from intentional canting is consistent with prior reports on two-fold symmetric superconductivity in NbSe$_2$, where the magnetoresistivity modulation is explained by $sin[2(\varphi + \varphi_0)]$ in the absence of canting and $|sin(\varphi + \varphi_0)|$ in the presence of canting[55]. Thus, these measurements, illustrating the stark contrast between cases with and without intentional canting, allow us to rule out the effects of accidental canting in our upper critical field measurements presented in Figs. 1**e** and 1**f**.

We would also like to address the persistent two-fold modulation observed in the normal state, indicating the presence of two-fold symmetry in the normal state of $CsV_3Sb_5$. The normal state of $CsV_3Sb_5$ is characterized by a charge density wave, which, as reported in previous experiments[15], leads to a nematic state and results in a two-fold symmetric oscillation in the magnetoresistivity as a function of $\varphi$. In Extended Fig. 4**a**, we present angular-dependent magnetoresistivity measurements at $\theta = 90^0$ in the normal state ($T = 5.0$ K). We observe a clear two-fold symmetric modulation that becomes more prominent with higher magnetic fields, which is consistent with the increase in magnetoresistivity. This trend contrasts with the two-fold symmetric modulation depicted in Extended Figs. 3**a** and 3**b**, where it is more pronounced within the field range encompassing the onset and offset of the superconducting state.

### VII.     Discussion on the in-plane angle dependency of $H_{c2}$

The flakes used in the devices for electrical transport measurements, such as Sample I, appear rectangular; see the inset of Fig. 1**b** for the optical microscopy image of Sample I. We estimate the rectangular dimensions of Sample I to be approximately: $a \sim 36$ $\mu$m, $b \sim 14$ $\mu$m, $c \sim 90$ nm, leading to a diagonal dimension $d \sim 38.6$ $\mu$m. According to ref.[56], the demagnetization factor $N$ for a cuboid is given by the following expression:

$$N^{-1} = 1 + \frac{3}{4}\frac{c}{a}\left(1 + \frac{a}{b}\right)$$

Here, $c/a$ = 90 x 10$^{-9}$ / 36 x 10$^{-6}$ = 0.0025 and $a/b$ = 36/14 = 2.57, and field is assumed to be applied along the *c*-axis. Therefore:

$$N^{-1} = 1 + \frac{3}{4} \times 2.5 \times 10^{-3} \times (1 + 2.57)$$

$N = (1.0067)^{-1} = 0.993$. This value is very close to 1, which is the well-known value for a thin sheet of magnetic material with the magnetic field applied perpendicularly to its plane. Given the very low $c/a$ or $c/b$ ratios, our sample can also be approximated to a thin sheet of magnetic material, for which the demagnetization factor for fields applied along a planar direction can be approximated to $N \approx 0$.

Quantitatively, using the formula given in ref.[57], we can estimate the demagnetizing factor for in-plane fields along the two edges of our sample. With dimensions of approximately 36 $\mu$m in length, 14 $\mu$m in width, and 90 nm in thickness, we obtain $N \approx 0.0126$ for the field along the width and $N \approx 0.0048$ for the field along the length. In this approximation, the demagnetization factor varies from 0.5% to 1.2% when the field is rotated within the planes from one edge to another at 90 degrees. These values are significantly smaller than the 8% variation in the angular dependency of $\mu_0 H_{c2}$ observed in Fig. 1**f**. Therefore, the demagnetization factor is very small and cannot account for the in-plane anisotropy observed in $H_{c2}$ (Fig. 1**f**).

## VIII. Thermal transport measurements

The thermal transport measurements were conducted at the National High Magnetic Field Laboratory in Tallahassee, Florida, USA, utilizing an 18 T magnet equipped with a $^3$He system. To measure thermal conductivity, we employed a conventional one-heater three-thermometer method. Accurate temperature sensing was achieved using field calibrated Cernox temperature sensors. These sensors were carefully glued onto an exfoliated sample, typically having dimensions of 3 x 1.5 x 0.07 mm$^3$, using silver paste. In the experiment, a heat pulse was applied to induce a longitudinal thermal gradient corresponding to 1-2% of the sample's base temperature. After applying the heat pulse, the temperatures of all three thermometers were continuously monitored until they reached a stable condition.

## IX. Magnetic field dependence of the thermal conductivity

In the main text, we presented data on the temperature-dependent thermal conductivity ($\kappa_{xx}/T$) to investigate the crossover between regimes I and II, as well as the angular variation of $\kappa_{xx}/T$ to highlight its two-fold symmetry. In this section, we provide additional measurements of thermal conductivity as a function of magnetic field, which are summarized in Extended Fig. 6, to capture the key features of the two superconducting regimes.

We begin by presenting the field dependence of $\kappa_{xx}/T$ for magnetic fields directed along the *ab* plane (Extended Fig. 6**a,b**) at two temperatures, $T \simeq 0.7$ K (regime II, panel **a**) and 1.1 K (regime I, panel **b**). In both cases, the data clearly shows a sharp increase in $\kappa_{xx}/T$ as the magnetic field is swept from zero. As the field approaches $H \simeq H_{c2}$, the slope of $\kappa_{xx}/T(H)$ decreases sharply, manifesting what seemingly are distinct superconducting transitions in both regimes, with the field positions aligning with the results obtained from electrical transport measurements.

The field dependence of $\kappa_{xx}/T$ provides important insights into the superconducting gap function. In nodeless (such as *s*-wave) superconductors, $\kappa_{xx}/T$ shows exponential growth with a very slow increase in $H$ at low fields[40,58]. By contrast, in nodal superconductors (such as *d* or *p*-wave), the quasiparticle conduction and $\kappa_{xx}/T$ increase rapidly once the field surpasses $H_{c1}$ (which is typically very small)[40,59]. In both Extended Figs. 6**a** and 6**b**, we observe a steep increase in $\kappa_{xx}/T$ at low fields. This sharp increase, contrasting with the exponentially slow growth seen in *s*-wave superconductors at $H \ll H_{c2}$, suggests that the thermal transport in CsV$_3$Sb$_5$ might be governed by delocalized quasiparticles arising from possible gap nodes.

## X. Heat capacity measurements

A differential, SiN membrane-based nanocalorimeter was used for measuring the specific heat of CsSb$_3$V$_5$ crystals. This technique is suitable for studying samples having very small masses. The measurements were performed in a dilution refrigerator coupled to a superconducting magnet capable of reaching a maximum field µ$_0$H = 16 T. As described in ref.[60], the calorimeter consists of a pair of cells, each composed of a stack of heaters and a thermometer at the center of the SiN membrane, yielding a background heat capacity inferior to 100 nJ/K at 300 K, and to 10 pJ/K at 1 K.

The sample is placed directly in contact to the resistive thermometer, which is made of a Ge$_{1-x}$Au$_x$ alloy, and displays a high dimensionless sensitivity $|dlnR/dlnT| \gtrsim 1$ over the entire temperature range. The data were acquired with a set of eight synchronized lock-in amplifiers measuring the DC, 1st, and 2nd harmonic signals of heaters and thermometers. Absolute accuracy is achieved *via* a variable-frequency-fixed-phase technique with the measurement frequency automatically adjusted during the data acquisition to account for the temperature changes associated to the sample heat capacity and the thermal conductance of the calorimeter.

### XI. Anisotropy in angular thermal conductivity

Here, we briefly outline the factors that could contribute to the anisotropy in angular thermal conductivity observed in our data. As discussed in the main text, three possible sources of anisotropy for $k_{xx}/T$ are identified:

1. At a fixed field, the anisotropy of the upper critical field ($H_{c2}$) results in a field-direction-dependent critical temperature. When combined with a strong temperature dependence of $k_{xx}/T$, this results in an anisotropy with a maximum in $k_{xx}/T$ along the direction of the highest $H_{c2}$.
2. In the mixed state, a two-fold symmetric signal is expected, with a maximum for fields aligned with the transport direction.
3. $k_{xx}/T$ can also be angle-sensitive for (near) nodal superconducting gap structures. This sensitivity arises due to either the density of heat-carrying quasiparticles and the Doppler shift of quasiparticle excitations outside of vortex cores, known as the Volovik effect.

Given these possibilities, our data suggest a strongly anisotropic or even nodal gap on one of the pockets for the following reasons: At higher temperatures, the anisotropy is likely dominated by the anisotropy observed in $\mu_0 H_{c2}$. The anisotropy of $\mu_0 H_{c2}$ does not rotate as a function of the temperature down to our lowest temperatures. Moreover, since the direction of the upper-critical-field anisotropy does not change between the two regimes (see Extended Fig. 7 and Methods Section XII), sources (1) and (2) cannot explain the observed anisotropy. Regarding source (3), we attribute the anisotropy at lower temperatures to the Doppler shift of the nodal quasiparticles and the associated increase in the density of states. This conclusion is supported by the fact that, for the temperatures considered, the thermal conductivity increases as a function of the field, as shown in Extended Fig. 6 (see ref. [40] for a detailed discussion of the various factors contributing to the thermal conductivity). This implies that the anisotropy stems from the Volovik effect, indicating anisotropy in one of the gap functions.

### XII. (In-plane) angular variation of the upper critical field at $T = 1.8$ K

In this section, we investigate the angular dependence of the upper critical field at $T = 1.8$ K as we vary the direction of the magnetic field along the *ab*-plane. In the main text, we presented similar data taken at $T = 0.3$ K, which falls within regime II (Fig. 1**e**, **f**) and exhibits a clear two-fold symmetry in $\mu_0 H_{c2}$. Here, at $T = 1.8$ K, we aim to determine whether this two-fold symmetry persists within regime I. We summarize our findings in Extended Fig. 7. In Extended Fig. 7**a**, we plot the magnetoresistivity for different in-plane rotation angles, $\varphi$, and observe a progressive shift in the superconducting transition along the magnetic field axis, resulting in an anisotropic $\mu_0 H_{c2}$, akin to the behavior observed in Fig. 1**e**. It is worth noting that, here, $\varphi = 0°$ is defined as the in-plane angle where $H_{c2}$ reaches its maximum value. Furthermore, consistent with the Fig. 1**e** data, the polar plot of $H_{c2}$ as a function of $\varphi$ reveals the presence of an emergent two-fold symmetry (Extended Fig. 7**b**). This observation indicates that the two-fold symmetry persists within regime I as well.

### XIII. Angular dependence of thermal conductivity

In Extended Fig. 9, we show the angular ($\phi$)-dependent *ab*-plane thermal conductivity at 1.1 K (regime I) and 0.7 K (regime II) as the field is rotated within the *ab*-plane, specifically at 4 T (additional data at different field values are provided in Fig. 4), which is below $H_{c2}$. Both traces exhibit two-fold symmetric oscillations, consistent with the two-fold symmetry observed in the electrical transport experiments shown in Fig. 1**e**,**f**. It is worth noting that due to the heat flow direction along the *ab* plane, a two-fold symmetric signal is expected, with a maximum at the direction of heat flow (defined as $\varphi = 0$) and a minimum perpendicular to it ($\varphi = 90^o$), resulting from the difference in heat transport along and normal to the vortices[40,61]. Upon closer examination, however, we find that the maximum for 1.1 K and 0.7 K does not align with $\varphi = 0$ in either set of data. Moreover, there is an apparent $\pi$ phase shift between the two $\kappa_{xx}/T$ traces at the two temperatures, where the maxima of one trace align with the minima of the other trace. To quantitatively analyze the phase shift, we fit the oscillatory component of the two data sets using the function $\cos 2\phi + \cos 2(\phi + \phi_0)$. The $\cos 2\phi$ term accounts for the two-fold symmetry arising from the difference in heat transport along and normal to the vortices, while the $\cos 2(\phi + \phi_0)$ term captures the two-fold symmetric term corresponding to the two-fold symmetric superconducting state inferred from our electrical transport experiments. Both the 1.1 K and 0.7 K data can be accurately fitted with this function, allowing us to extract $\phi_0$ as $-86.4 \pm 0.4^o$ and $88.7 \pm 1.3^o$ for 1.1 K and 0.7 K, respectively. Consequently, a phase shift of $175.1 \pm 1.7^o$ is obtained, which is remarkably close to $\pi$.

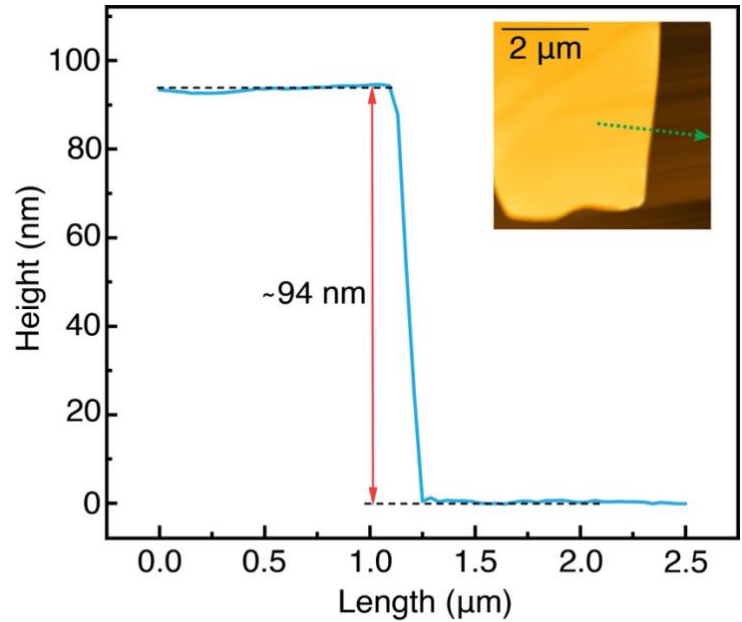

**Extended Fig. 1: Determination of the CsV$_3$Sb$_5$ flake thickness using atomic force microscopy.** Inset: Atomic force microscopy image of a CsV$_3$Sb$_5$ flake with similar optical contrast as the one used to collect the data in Fig. 1**b**. The height profile is plotted along the green dashed line (arrow indicates the scan direction) in the image, revealing a thickness of approximately 94 nm. This relatively thick flake is expected to exhibit properties that are representative of bulk CsV$_3$Sb$_5$.

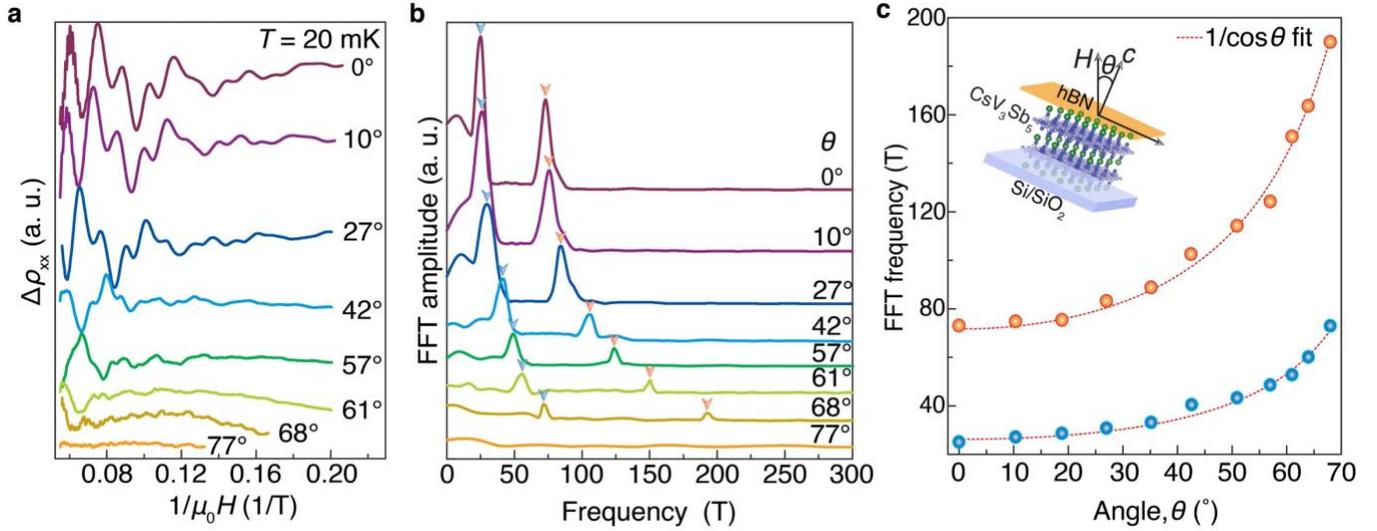

**Extended Fig. 2: Angular dependence of the Shubnikov de Haas oscillations in $CsV_3Sb_5$. A**, Oscillatory component superimposed onto the magnetoresistivity ($\Delta\rho_{xx}$) traces shown in Fig. 1**c**. These traces were obtained after subtracting a smooth background and recorded at different angles ($\theta$) between the magnetic field and the crystallographic *c*-axis. $\Delta\rho_{xx}$ is plotted as a function of $1/\mu_0 H$ to reveal the $1/\mu_0 H$-periodic quantum oscillations. Traces are vertically offset for clarity, and the corresponding $\theta$ values are provided for each trace. **b**, Fourier transform spectra of the $\Delta\rho_{xx}$ traces shown in panel **a**. Two discernible peaks appear in the spectra for each $\theta$ angle when $\theta \leq 68^o$. However, for $\theta$ values above $68^o$, no clear oscillation is observed in $\Delta\rho_{xx}$, and consequently, no discernible Fourier transform peak is resolved. The peak positions, expressed in units of 1/T, represent the area of the Fermi surface in units of the cyclotron frequency. The peaks, highlighted by color-coded arrows, shift towards higher frequencies as $\theta$ increases. **c**, Angular dependence of the frequency positions of the Fourier transform peaks, obtained by tracking the two main peaks in the Fourier transform data at different angles. The inset illustrates the direction of rotation, i.e., the magnetic field is rotated from the *c*-axis towards the *ab*-plane. The frequencies are plotted as a function of $\theta$. Red dashed curves represent the $1/\cos\theta$ fits applied to the angular evolution of both peaks. The remarkable agreement between the data and the $1/\cos\theta$ dependence demonstrates the two-dimensional character of both Fermi surfaces.

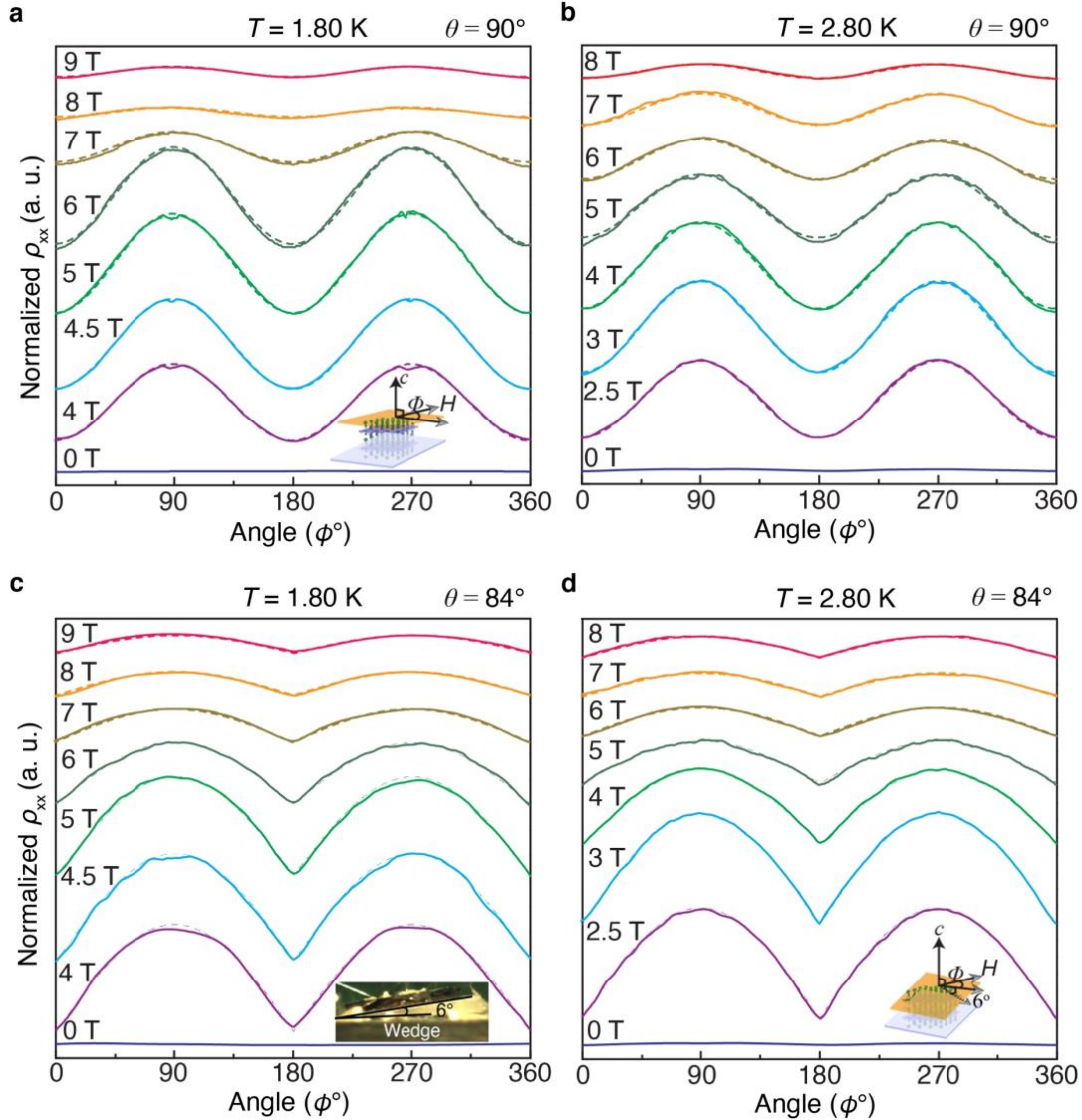

**Extended Fig. 3: Angular dependence of the magnetoresistivity along the *ab* plane obtained using the same sample under non-canted and canted conditions. a, b,** Evolution of the magnetoresistivity of the CsV$_3$Sb$_5$ device in a magnetic field rotating within the *ab*-plane as a function of $\varphi$ without a canting angle (i.e., at $\theta = 90^0$). Dashed lines represent fits to $sin[2(\varphi + \varphi_0)]$. Two sets of data are presented, collected at two different temperatures, $T = 1.8$ K (panel **a**) and 2.8 K (panel **b**), both below $T_c$. A clear two-fold symmetric pattern consistent with the results shown in Fig. 1**f** is observed. The two-fold symmetric oscillation is particularly prominent within the field range encompassing the onset and offset of the superconducting state. **c, d,** Angular dependence of the magnetoresistivity of the same sample collected with a canting angle of 6° (i.e., at $\theta = 84^0$). The photograph in the inset of panel **d** displays the sample positioned on a wedge to achieve a canting angle of 6°. Dashed lines denote fits to $|sin(\varphi + \varphi_0)|$. The data are shown for two different temperatures, $T \simeq 1.8$ K (panel **c**) and 2.8 K (panel **d**), both below $T_c$. In panels **a-d,** each trace is normalized by its maximum value to emphasize the strength of the two-fold symmetric oscillation. To enhance clarity, the traces corresponding to different magnetic fields at a specific temperature are vertically offset. The insets in panels **a** and **d** illustrate the corresponding directions of rotation.

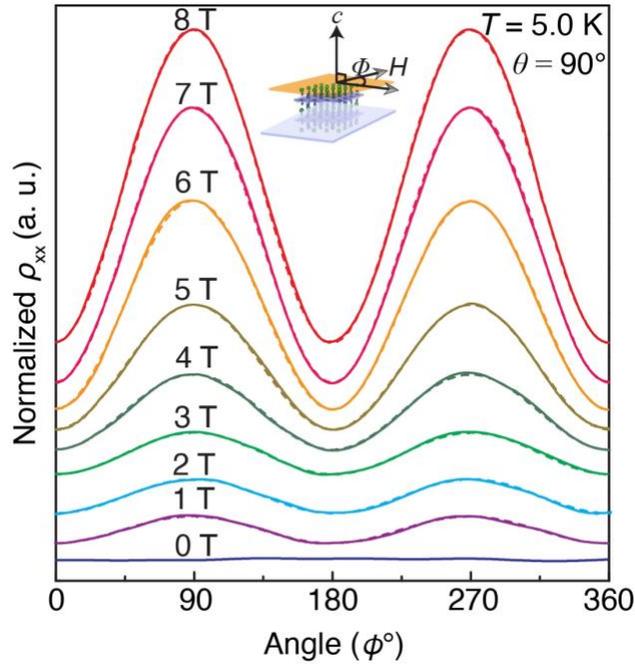

**Extended Fig. 4: Angular dependence of the magnetoresistivity in the normal state**. Evolution of the magnetoresistivity of $CsV_3Sb_5$ under a magnetic field rotating within the *ab*-plane (at $\theta = 90^0$) as a function of $\varphi$. The data were collected at $T \simeq 5.0$ K, or above $T_c$. Traces corresponding to different magnetic fields are vertically offset for clarity, and each trace is normalized by its maximum value to highlight the strength of the magnetoresistivity oscillation. Dashed lines represent $sin[2(\varphi + \varphi_0)]$ fits. The observed pattern aligns with the findings in Extended Fig. 3**a**, **b**, demonstrating a clear two-fold symmetric pattern.

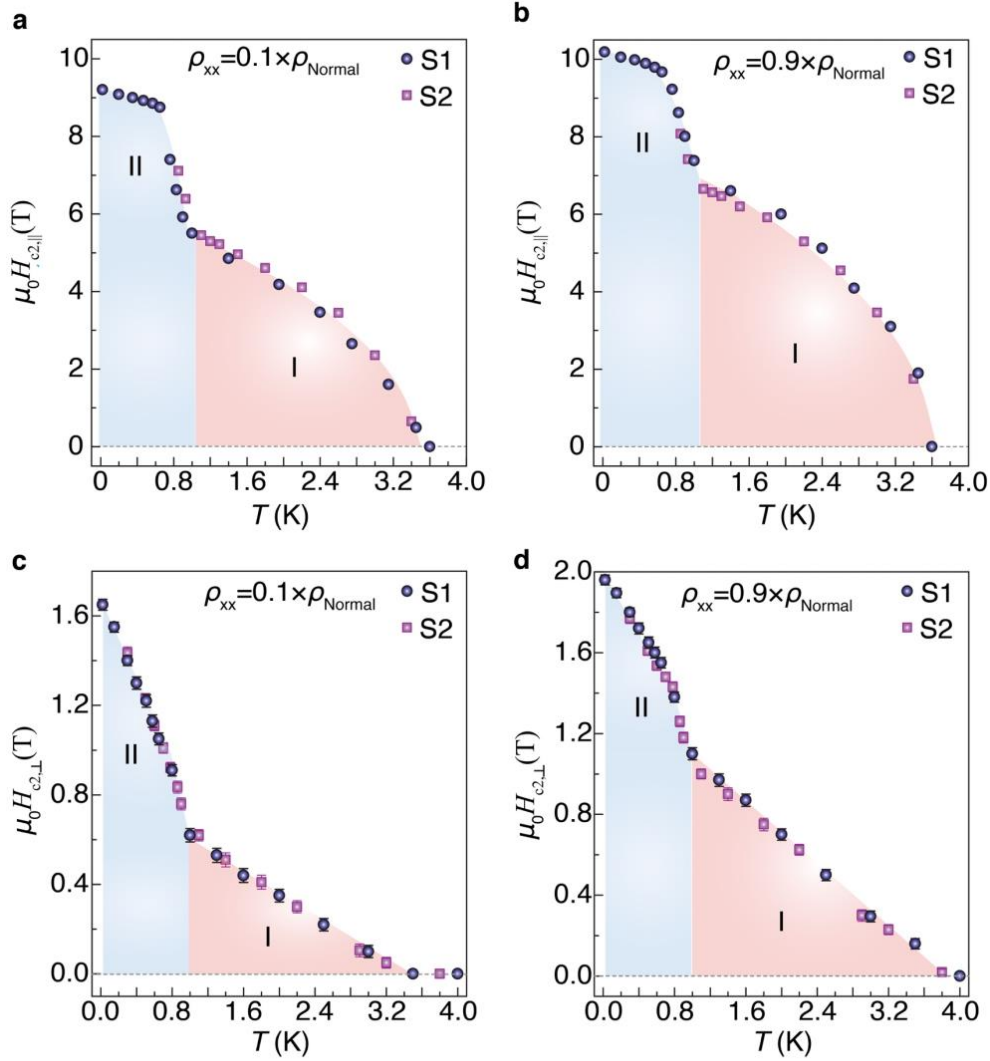

**Extended Fig. 5: Superconducting phase diagrams extracted by using the 10% and 90% of the resistivity in the normal state criteria, revealing two regimes. a**, **b**, In-plane upper critical field $\mu_0 H_{c2,\parallel}$, extracted from the traces in Figs. 2**a** and 2**b**, as a function of the temperature (*T*). The plot reveals two clear regimes (I and II) comprising distinct temperature regions. Data from two samples having similar thicknesses, S1 and S2, reveal overlapping phase-diagrams. **c, d**, Out-of-plane upper critical field $\mu_0 H_{c2,\perp}$, extracted from the traces in Figs. 2**d** and 2**e**, as a function of the *T*. This phase diagram also features both regimes (I and II) displayed in panels **a, b**. In panels **a** and **c**, $\mu_0 H_{c2}$ is defined according to $\rho_{H_{c2}} = 0.1 \times \rho_{\text{Normal}}$, whereas in panels **b** and **d**, $\mu_0 H_{c2}$ is defined using $\rho_{H_{c2}} = 0.9 \times \rho_{\text{Normal}}$, where $\rho_{\text{Normal}}$ is the normal state resistivity just above the superconducting transition. To extract $\rho_{\text{Normal}}$, we fitted $\rho_{xx}$ to a straight line, subsequently choosing its deviation from linearity as the value of $\rho_{\text{Normal}}$. The error bar is given by the uncertainty on the exact field value where the deviation from linearity occurs. For $\mu_0 H_{c2,\parallel}$, the error bar is ~±0.01 T, which is smaller than the symbol size in panels **a** and **b**. For $\mu_0 H_{c2,\perp}$, the error bar is ~±0.03 T, matching the symbol size in panels **c** and **d**. Hence, we included error bars in panels **c** and **d**. While the differences between phase-diagrams drawn using the 90% and 10 % criteria can be attributed to vortex physics, the phase diagrams, importantly, show the same features pointing to two distinct superconducting regimes.

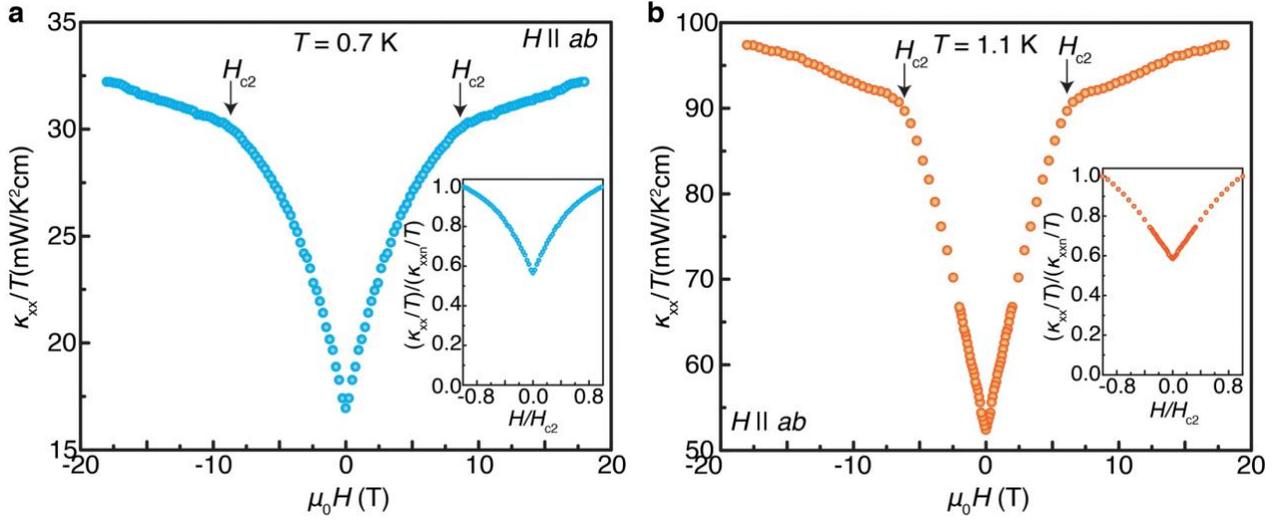

**Extended Fig. 6: Thermal conductivity in the two superconducting regimes (below and above $T \simeq 0.8$ K). a, b,** Field dependence of the *ab*-plane thermal conductivity at temperatures $T \simeq 0.7$ K (panel **a**) and $T \simeq 1.1$ K (panel **b**) with a field oriented along the *ab*-plane at $\varphi = 0$. The positions of $H_{c2}$ (obtained from electrical transport measurements at the respective temperatures) are indicated. As the magnetic field is increased from zero, thermal conductivity rises rapidly until $\mu_0 H \simeq \mu_0 H_{c2}$. The insets show the corresponding $\kappa_{xx}/T$ normalized by its normal state value and plotted against the magnetic field normalized by $H_{c2}$. At $T = 0.7$ K, $\kappa_{xx}/T$ does not follow a well-defined power law, e.g., linear in field, or increases exponentially as seen in multiband and conventional *s*-wave superconductors, respectively. $\kappa_{xx}/T$ does display a linear in $\mu_0 H$ dependence at $T = 1.1$ K.

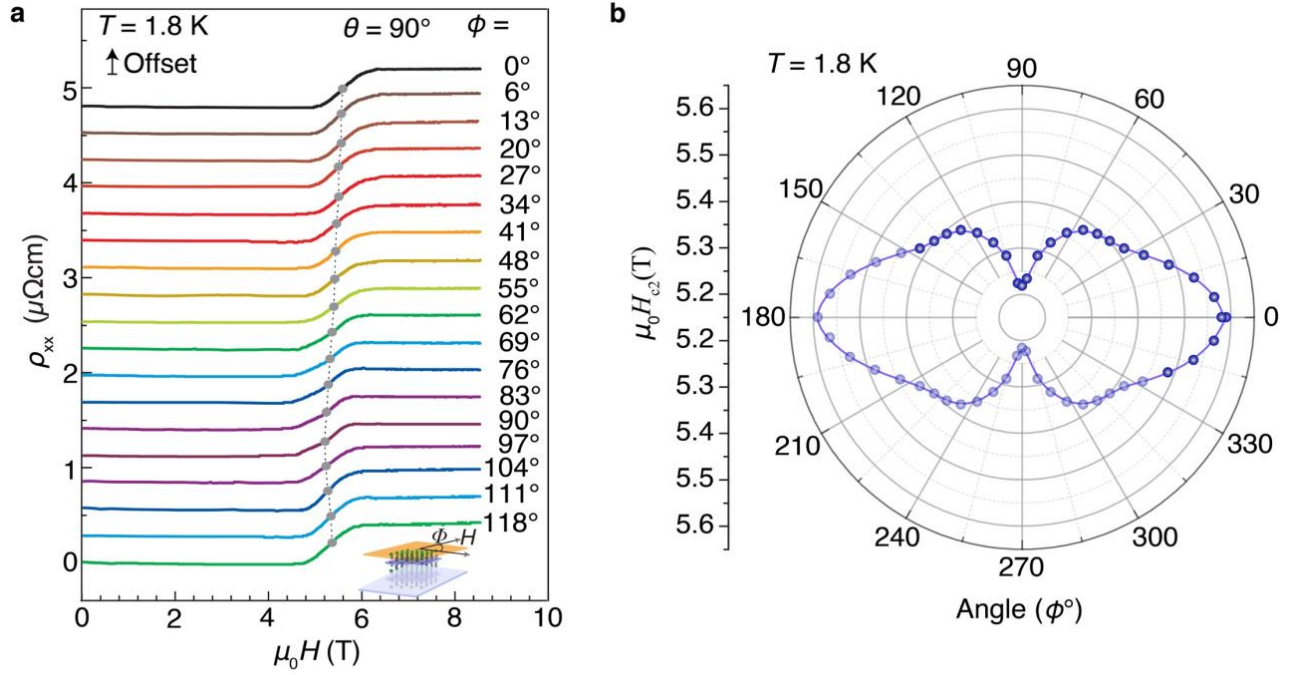

**Extended Fig. 7: Two-fold symmetric superconductivity in CsV$_3$Sb$_5$ at $T \simeq 1.8$ K. a,** Magnetic field dependence of the resistivity at $T \simeq 1.8$ K with various in-plane rotation angles ($\varphi$). The inset illustrates the direction of rotation, where $\varphi$ represents the angle between the in-plane magnetic field (along the *ab* plane) and the axis with the highest in-plane upper critical field. The upper critical magnetic field is indicated by grey markers on each trace. Traces corresponding to different $\varphi$ angles are vertically offset for clarity. **b,** $\varphi$-dependence of in-plane upper critical field, revealing a distinct two-fold symmetry along the *ab* plane, akin to the Fig. 1**e** data obtained at $T \simeq 0.3$ K. The darker (lighter) color symbols in panels **a** and **b** denote raw (repeated assuming 180° periodicity) data.

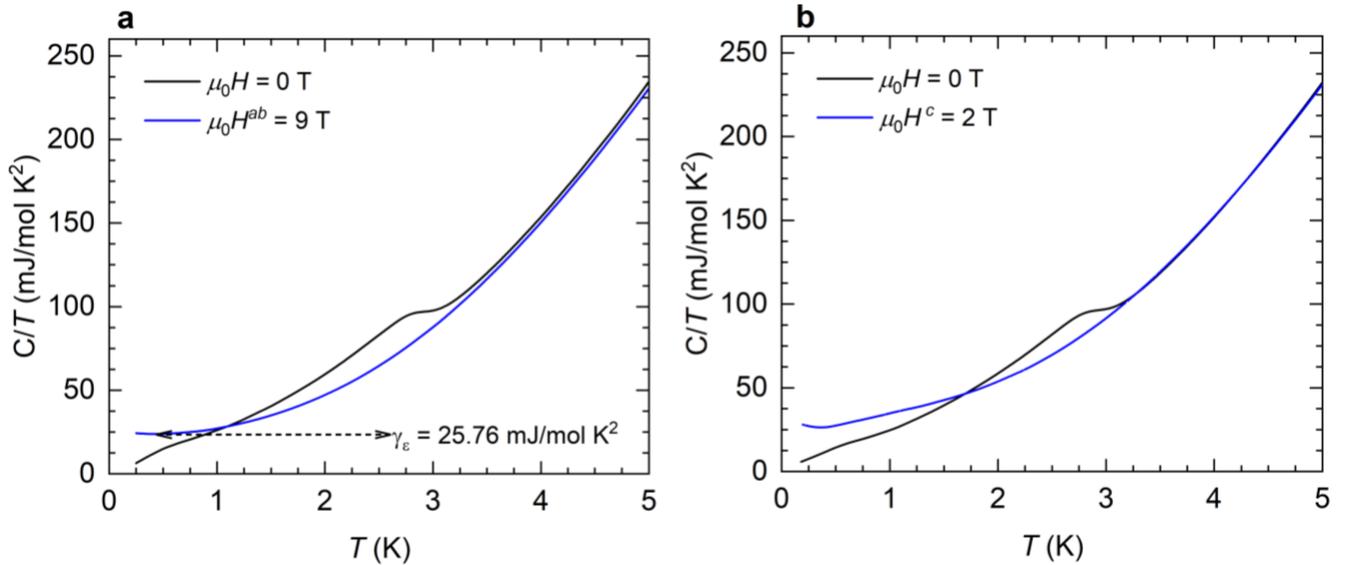

**Extended Fig. 8: Raw heat capacity data as a function of the temperature and for two field orientations. a,** $C/T$ as a function of $T$ for $\mu_0 H = 0$ T (black trace) and 9 T (blue trace) applied along the *ab*-plane, respectively. **b,** $C/T$ as a function of $T$ for for $\mu_0 H = 0$ T (black trace) and 2 T (blue trace) applied along the $c$-axis, respectively. The traces measured under an external field were subtracted from the zero field one, to yield the electronic contribution shown in Fig. 3c within the main text.

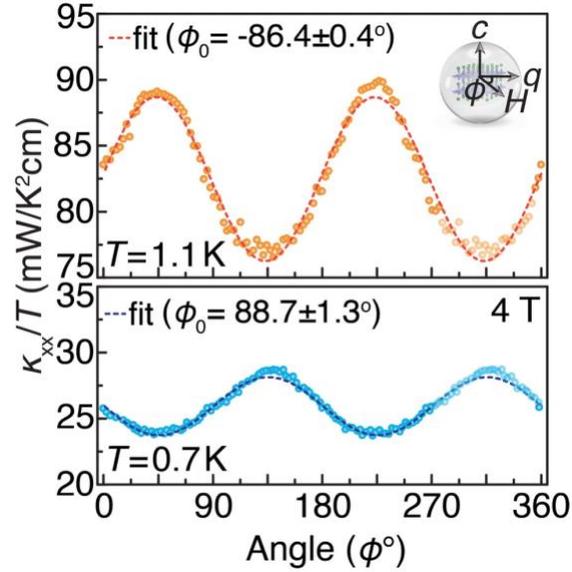

**Extended Fig. 9: Evidence of two-fold symmetric superconductivity in thermal transport experiments.** Angular dependence of the in-plane thermal conductivity at a fixed magnetic field of $\mu_0 H = 4$ T, as the field is rotated within the $ab$-plane. The measurements were performed at temperatures $T \simeq 0.7$ K (bottom panel) and $T \simeq 1.1$ K (top panel). The angle $\phi$ represents the orientation between the in-plane magnetic field and the direction of the applied heat current ($q$), as illustrated in the inset of the top panel. Both temperature datasets display distinct two-fold symmetric oscillations. The oscillatory component of the data can be accurately fitted to the function $\cos 2\phi + \cos 2(\phi + \phi_0)$. The color-coded dashed curves represent the fitted curves, and the corresponding fitted $\phi_0$ values are provided for each panel. Additionally, the fitting reveals an apparent $\sim \pi$ phase shift between the 0.7 K and 1.1 K data, as visually evident from the data. The darker (lighter) color symbols denote raw (repeated assuming 180° periodicity) data.

**Competing interests:** The authors declare no competing interests.

**Data and materials availability:** All data needed to evaluate the conclusions in the paper are present in the paper. Additional data are available from the corresponding authors upon reasonable request.

**Acknowledgement:** M.Z.H. group acknowledges primary support from the US Department of Energy, Office of Science, National Quantum Information Science Research Centers, Quantum Science Center (at ORNL) and Princeton University; STM Instrumentation support from the Gordon and Betty Moore Foundation (GBMF9461) and the theory work; and support from the US DOE under the Basic Energy Sciences programme (grant number DOE/BES DE-FG-02-05ER46200) for the theory and sample characterization work including ARPES. The sample growth was supported by the National Key Research and Development Program of China (grant nos 2020YFA0308800 and 2022YFA1403400), the National Science Foundation of China (grant no 92065109), and the Beijing Natural Science Foundation (grant nos Z210006 and Z190006). Z.W. thanks the Analysis and Testing Center at BIT for assistance in facility support. L.B. is supported by DOE-BES through award DE-SC0002613. The National High Magnetic Field Laboratory (NHMFL) acknowledges support from the US-NSF Cooperative agreement Grant number DMR-DMR-2128556 and the state of Florida. We thank T. Murphy, G. Jones, L. Jiao, and R. Nowell at NHMFL for technical support. T.N. acknowledges supports from the European Union's Horizon 2020 research and innovation programme (ERC-StG-Neupert-757867-PARATOP). Z.W. thanks the Analysis and Testing Center at BIT for assistance in facility support.